\documentclass[11pt]{article}
\usepackage{times}
\usepackage{geometry}
\usepackage[utf8]{inputenc}
\usepackage{enumitem,amssymb}
\usepackage{ragged2e}
\newlist{thematic}{itemize}{8}
\setlist[thematic]{label=$\square$}
\usepackage{pifont}
\usepackage{fancyhdr}
\usepackage[USenglish]{babel}
\usepackage[utf8]{inputenc}
\usepackage[T1]{fontenc}
\usepackage{epsfig}
\usepackage{wrapfig}
\usepackage{color}

\usepackage[vflt]{floatflt}
\usepackage{longtable}
\usepackage{caption}
\usepackage{jheppub}   
\usepackage[dvipsnames,svgnames,x11names]{xcolor}
\usepackage[all]{hypcap} %Links go to figures; breaks on deluxetables (use \capstartfalse \capstarttrue to fix it)
\usepackage{amssymb,amsmath}
\usepackage{longtable}
\usepackage{amssymb}
\usepackage{amsmath, xspace}
\usepackage{graphics,graphicx} 
\usepackage{rotating}
\usepackage{color}
\usepackage{xcolor}
\usepackage{multirow}
\usepackage{subfigure}
\usepackage{sectsty}
\usepackage[outercaption]{sidecap}
\sidecaptionvpos{figure}{c}

\usepackage{enumitem}
\setlist[enumerate]{itemsep=0pt, parsep=0pt}
\setlist[itemize]{itemsep=0pt, parsep=0pt}

\definecolor{DarkGreen}{rgb}{0.0, 0.3, 0.0}
\definecolor{purple}{rgb}{0.5, 0.0, 0.5}
\definecolor{red}{rgb}{1, 0.0, 0.0}
\definecolor{green}{rgb}{0, 1.0, 0.0}

\def\3he{$^3{\rm He}$}

\def\lsim{\mathrel{\lower2.5pt\vbox{\lineskip=0pt\baselineskip=0pt
           \hbox{$<$}\hbox{$\sim$}}}}

\def\gsim{\mathrel{\lower2.5pt\vbox{\lineskip=0pt\baselineskip=0pt
           \hbox{$>$}\hbox{$\sim$}}}}

\begin{document}
\raggedright
\huge
Surveying Galaxy Clusters (in formation) in the Distant Universe 
\linebreak
\bigskip
\normalsize

\textbf{Authors:} 
J.M. Pérez-Martínez$^{1,2}$ (jm.perez@iac.es), H. Dannerbauer$^{1,2}$, E. van Kampen$^{3}$, C. Cicone$^{4}$, E. Hatziminaoglou$^{3,1,2}$, L. Bonavera$^{5,6}$, L. Di Mascolo$^{7}$,  J. González‑Nuevo$^{5,6}$, F. Guglielmetti$^{3}$, A. Pensabene$^{8}$, J. van Marrewijk$^{9}$ \\

\bigskip
\textit{$^1$Instituto de Astrofísica de Canarias (IAC), E-38205, La Laguna, Tenerife, Spain \\
$^2$Universidad de La Laguna, Dpto. Astrofísica, E-38206, La Laguna, Tenerife, Spain \\
$^3$ESO, Karl-Schwarzschild-Str. 2, 85748 Garching bei München, Germany \\
$^4$Institute of Theoretical Astrophysics, University of Oslo, PO Box 1029, Blindern 0315, Oslo, Norway \\
$^5$ Departamento de Fisica, Universidad de Oviedo, C. Federico Garcia Lorca 18, 33007 Oviedo, Spain.\\
$^6$ Instituto Universitario de Ciencias y Tecnologías Espaciales de Asturias (ICTEA), C. Independencia 13, 33004 Oviedo, Spain.\\
$^7$Kapteyn Astronomical Institute, University of Groningen, Landleven 12, 9747 AD, Groningen, The Netherlands. \\
$^8$DAWN/DTU-Space, Electrovej 327, 2800 Lyngby, Denmark \\
$^9$Leiden Observatory, Leiden University, P.O. Box 9513, 2300, RA Leiden, The Netherlands. \\
}
\bigskip

\textbf{Science Keywords:} 
galaxies: clusters; galaxies: starburst; galaxies: high-redshift; galaxies: evolution; galaxies: formation; 

\setcounter{figure}{0}
\captionsetup{labelformat=default}

\pagebreak

\justifying
\section{Abstract}  
\vspace{-0.2cm}
Present-day galaxy clusters are the largest virialized structures in the Universe, yet their early assembly remains poorly understood. At z$>$2, clusters in formation span tens of Mpc and host gas-rich, dust-obscured galaxies embedded in extended, low–surface-brightness gaseous environments. Current (sub-)millimeter facilities lack the mapping speed, sensitivity, and contiguous field of view needed to trace the cold gas and dust driving rapid galaxy growth across such scales. A future large single-dish observatory with degree-scale coverage, broad spectral access, and high-multiplex capability would enable comprehensive and uniform mapping of entire protoclusters, revealing where star formation is triggered or quenched, and quantifying the cold gas budget, thus providing information on gas cooling within protocluster environments. In synergy with wide-sky optical/NIR surveys such as Euclid, LSST, and Roman, this facility would provide the missing multi-scale and multiphase submillimeter view needed to uncover how the stellar, gaseous, and dark-matter components assemble in protoclusters, completing our view of early structure formation.

\vspace{-0.35cm}
\section{Scientific context and motivation}
\vspace{-0.2cm}
Galaxy clusters in the local Universe are massive, virialized systems whose mass budget is dominated by hot gas and dark matter. Their progenitors, the so-called protoclusters at $z>2$, are instead diffuse, dynamically young structures extending over tens of megaparsecs (Overzier 2016; Alberts \& Noble 2022). These early environments host 20-50$\%$ of the star formation density of the Universe, display intense black-hole activity, and large-scale gas inflows and outflows, all embedded within the filamentary network of the cosmic web (e.g., Chiang et al. 2017; Daddi et al. 2021; Tozzi et al. 2022). Understanding how galaxies, gas, and dark matter assemble into the first cluster-scale halos is therefore essential for tracing the emergence of large-scale structure and the baryon cycle across cosmic time. Over the past two decades, submillimeter observations have been central to protocluster science (Casey 2016; Hodge \& da Cunha 2020). Wide far-infrared and (sub-)millimeter surveys with facilities such as Herschel, LABOCA, or Planck revealed statistically significant overdensities of dusty, gas-rich starbursts that marked the densest nodes of the cosmic web at the cosmic noon and beyond (e.g., Ivison et al. 2013; Dannerbauer et al. 2014; Bussmann et al. 2015; Planck Collaboration 2016; Negrello et al. 2017).

The advent of ALMA marked a transformational leap. Its unprecedented sensitivity and angular resolution enabled targeted observations of these candidate protoclusters, routinely resolving single-dish overdensities into individual member galaxies (e.g., Riechers et al. 2014; Oteo et al. 2018). Through detections of CO, [C\textsc{i}], [C\textsc{ii}], and dust continuum emission, prominent overdensities hosting extreme dust-obscured starburst activity and complex gas kinematics have been uncovered (e.g., Wang et al. 2016; Miller et al. 2018; Venkateshwaran et al. 2024). These works established ALMA as the primary tool for dissecting the internal physics of galaxies within nascent clusters. However, fundamental limitations persist. ALMA’s field of view is intrinsically small ($\sim$1'), restricting observations to tiny fractions of structures that span tens of megaparsecs. As a result, current interferometric mapping cannot track how star formation, gas accretion, or feedback propagate through the larger protocluster environment, nor can it recover the low–surface-brightness circumgalactic and intergalactic medium (CGM/IGM) that links galaxies to the surrounding cosmic web (Dannerbauer et al. 2017; Emonts et al. 2023; Harrington et al. 2025). Signatures of nascent fully ionized intracluster gas are even harder to detect: only few tentative cases exist, owing to their intrinsically weak contrast and the difficulty interferometers face in recovering extended emission (Di Mascolo et al. 2023; van Marrewijk et al. 2024; Zhou et al. 2025).

To overcome these challenges, we require a large-aperture, wide-field single-dish (sub)millimeter facility capable of mapping the full extent of forming clusters and their gaseous environments. Such an observatory must combine degree-scale coverage with broad spectral access similar to ALMA (30-950 GHz), high spectral resolution, and the mapping speed and surface-brightness sensitivity needed to recover both compact sources and diffuse emission. This would enable uniform, wide-field spectroscopic surveys of the cold baryonic content of protoclusters, detecting hundreds of member galaxies, extended CGM/IGM reservoirs, and potentially the earliest signatures of intracluster-medium formation. Crucially, this would provide the long-missing counterpart to wide-sky optical/NIR surveys such as Euclid, LSST, and Roman, and forthcoming highly multiplexed spectrographs (e.g., VLT/MOONS) tracing the stellar and warm ionized-gas components over vast volumes. Together, these datasets would deliver the first truly multi-phase, multi-scale view of forming clusters, revealing how the stellar, gaseous, and dark-matter components assemble and interact from galactic to cosmological scales.

\vspace{-0.35cm}
\section{Science case}
\vspace{-0.2cm}
Fundamental aspects of large-scale structure and galaxy co-evolution remain poorly constrained due to the lack of wide-field cold-gas sensitivity required to trace cluster formation across tens of Mpc. A next-generation single-dish submillimeter facility combining degree-scale mapping with multi-line spectroscopy would provide this missing perspective, enabling a contiguous view of the cold baryons during structure formation. The following science goals illustrate a new and unique discovery space:

\smallskip
\noindent
\textbf{\underline{When and how do the first galaxy clusters assemble on Mpc scales?:}}
Protoclusters at z$>$2 mark the earliest stages of cluster formation, yet the emergence of their large-scale structure, including the filaments, infalling groups, and diffuse gas reservoirs that extend up to 50 Mpc from the core, remains virtually unconstrained. Despite current single-dish facilities being able to probe such scales, their sensitivity limitations and coarse resolution hamper secure source identification, thus demanding complementary follow-up data (e.g., Bing et al. 2023; Lagache et al. 2025). On the other hand, deep interferometric observations can easily resolve the internal structure of compact protocluster cores (Long et al. 2020; Hill et al. 2020; Champagne et al. 2021; Jin et al. 2021), albeit their limited field of view per pointing ($\sim1-2'$, roughly 0.5-1 Mpc) is far too small to capture the full network of gas and galaxies assembling the protocluster. As a result, we lack any contiguous view of how cold gas is arranged across the surrounding filaments and infalling substructures, or how these components connect dynamically to the forming cluster core. A wide-field single-dish submillimeter facility would transform this picture by mapping the cold gas (e.g., CO and Carbon lines) and dust emission across entire protoclusters in a single visit, tracing the distribution and kinematics of cold gas from the field down to the cluster core. Such observations would reveal which assembly channels dominate, including steady filamentary accretion, mergers of gas-rich groups, or more rapid top-down collapse, and how these processes evolve across cosmic time. By providing complete, uniform maps of full protoclusters and enabling such studies for large samples, this facility would allow empirical discrimination between competing theoretical models and establish a statistically robust framework for understanding the earliest phases of cluster growth.

\smallskip
\noindent
\textbf{\underline{How do protocluster environments on scales of 50 to 500 kpc regulate galaxy growth?:}} Protocluster galaxies span a wide range of evolutionary stages, yet many display elevated AGN activity, episodes of intense star formation, or unusually extended molecular gas reservoirs relative to field galaxies (Calvi et al. 2023; Chen et al. 2024; Vito et al. 2024).
Furthermore, the triggering of AGN and the impact of their outflows on heating or disturbing the surrounding gas remain difficult to assess (e.g., Chapman et al. 2025). The crucial missing information lies in the diffuse circumgalactic and intergalactic medium that supplies, regulates, and removes gas from these galaxies. Current interferometers filter out this extended, low surface brightness emission, leaving key components of the cold gas reservoir, including filamentary bridges and diffuse halos, largely undetected. Early signatures of ICM formation, such as weak SZ signals, have been observed in only a few extreme systems (Di Mascolo et al. 2023), underscoring the sensitivity and spatial-scale limitations of present facilities. A wide-field single-dish submillimeter observatory capable of recovering such diffuse emission would reveal where and how gas cools and accretes onto galaxies, and if it is removed or heated by feedback-driven outflows or environmental interactions (e.g., tidal or ram pressure stripping or caused by nascent ICM), directly linking gas dynamics on intermediate scales to star formation and AGN activity during cluster assembly.

\smallskip
\noindent
\textbf{\underline{How do dense regions influence star formation on galactic scales?:}}
Protoclusters host a substantial fraction ($>$20\%) of the Universe’s star-formation rate density at $z>2$, yet the true distribution of this activity remains poorly constrained. Rest-frame optical surveys trace only the unobscured component of star formation and the buildup of stellar mass, missing the heavily dust-enshrouded activity (Zavala et al. 2021; Gottumukkala et al. 2024) that dominates many protocluster galaxies (e.g., Pensabene et al. 2024; Zhang et al. 2024). As a result, it remains unsettled how star formation is distributed between core and outskirts, how gas fractions and depletion times vary with environment, and how star-formation efficiency evolves as the structure grows (e.g., Wang et al. 2018; Pérez-Martínez et al. 2025). Existing and forthcoming wide-field surveys such as Euclid, LSST, and Roman will identify statistical protocluster samples and map their stellar and ionized gas content with rest-frame optical tracers, but they still lack the cold-gas diagnostics needed to recover the bulk of the obscured star formation and its environmental dependence. A new wide-field submillimeter facility would provide this missing gaseous component by delivering uniform multiline CO, [C\textsc{i}], [C\textsc{ii}], measurements for full protocluster populations (including assessing the CO SLED), thus efficiently securing redshifts and gas properties for dozens of individual members simultaneously. When combined with the stellar and ionized-gas maps from wide-field surveys, we will obtain the first complete, multi-phase census of protocluster galaxies. This synergy would determine the total contribution of nascent clusters to cosmic star formation and clarify how gas supply and local conditions regulate galaxy and SMBH growth during the peak epoch of structure formation.

\vspace{-0.35cm}
\section{Technical requirements}
\vspace{-0.2cm}

 A facility capable of addressing these science goals requires a large single-dish submillimeter telescope of $\sim$50 meters in diameter, delivering the sensitivity, spatial/angular resolution, and survey efficiency needed to probe, in a single observation, the full extent of forming clusters and their diffuse gas across the key redshift range $z=2$–7. Operating at a high, dry site similar to the ALMA plateau and covering a comparable frequency range (30-950 Ghz), such a telescope would achieve diffraction-limited performance of $\sim$2" at 700 GHz, reducing confusion noise and enabling precise measurements of CO, [C\textsc{i}], [C\textsc{ii}], and dust continuum from both compact galaxies and extended low–surface-brightness structures. An instantaneous field of view close to 2 degrees in diameter ($\sim50\times50$ Mpc at $z>2$) would provide true wide-field access to protocluster environments, combining unparalleled mapping speed with sensitivity to the faint, diffuse emission that interferometers inherently resolve out. This wide-field, high-throughput mapping capability lies at the core of the Atacama Large Aperture Submillimetre Telescope (AtLAST) concept (Mroczkowski et al. 2025), which is specifically designed to deliver the sensitivity and surface-brightness recovery required for such transformative observations. With survey speeds 3-5 orders of magnitude faster than ALMA and continuum sensitivities comparable to its full array, AtLAST would supply the large-scale environmental context inaccessible to current interferometric observations. A flexible instrument suite enabling simultaneous multiband spectroscopy and high-fidelity continuum imaging would deliver comprehensive data products, including 3D spectral cubes, deep continuum maps, line-luminosity and gas-mass catalogs, and spatially resolved environmental-density fields, capturing the full cold baryonic content of forming clusters. Together, these elements define a transformative submillimeter capability that complements and completes ALMA’s interferometric strengths, enabling multi-scale exploration of the cold baryon cycle and structure formation in the early Universe.
\\

\vspace{-0.2cm}
{\sc{\textbf{References:}}}
Alberts et al. 2022, Univ, 8, 554 $\bullet$
Bing et al. 2023, A\&A, 677, 66 $\bullet$
Calvi et al. 2023, A\&A, 678, 15 $\bullet$
Casey 2016, ApJ, 824, 36 $\bullet$
Champagne et al. 2021, ApJ, 913, 110 $\bullet$
Chapman et al. 2025, arXiv:2511.17814 $\bullet$
Chen et al. 2024, MNRAS, 527, 8950 $\bullet$
Chiang et al. 2017, ApJL, 844, L23 $\bullet$
Dannerbauer et al. 2014, A\&A, 570, 55 $\bullet$
Dannerbauer et al. 2017, A\&A, 608, 48 $\bullet$
Di Mascolo et al. 2023, Nature, 615, 809 $\bullet$
Emonts et al. 2023, Sci, 379, 6639 $\bullet$
Gottumukkala et al. 2024, MNRAS, 530, 1 $\bullet$
Harrington et al. 2025, A\&A, 701, 298 $\bullet$
Hill et al. 2020, MNRAS, 495, 3, 3124 $\bullet$
Jin et al. 2021, A\&A, 652, 11 $\bullet$
Lagache et al. 2025, arXiv:2506.15322 $\bullet$
Long et al. 2020, ApJ, 898, 133 $\bullet$
Miller et al. 2018, Natur, 556, 7702 $\bullet$
Mroczkowski et al. 2025, A\&A, 694, 142 $\bullet$
Negrello et al. 2017, MNRAS, 470, 2253 $\bullet$
Oteo et al. 2018, ApJ, 856, 72 $\bullet$
Overzier 2016, A\&ARv, 24, 1 $\bullet$
Pensabene et al. 2024, A\&A, 684, 119 $\bullet$
Pérez-Martínez et al. 2025, A\&A, 696, 236 $\bullet$
Planck Collaboration. 2016, A\&A, 596, 100 $\bullet$
Remus et al. 2023, ApJ, 950, 191 $\bullet$
Smail et al. 2024, MNRAS, 529, 2290 $\bullet$
Venkateshwaran et al. 2024, ApJ, 977, 161 $\bullet$
Vito et al. 2024, A\&A, 689, 130 $\bullet$
Wang et al. 2018, ApJL, 867, L29 $\bullet$
Zavala et al. 2021, ApJ, 909, 165 $\bullet$
Zhang et al. 2024, A\&A, 692, 22 $\bullet$
Zhou et al. 2025, A\&A, 701, 234 $\bullet$
Zhou et al. 2025, arXiv:2509.03912 %$\bullet$

\end{document}